\documentclass[%
 pra,
 amsmath,amssymb,
 reprint,%
]{revtex4-1}

\usepackage{amsmath}
\usepackage{amssymb}
\usepackage{float}
\usepackage[utf8]{inputenc}
\usepackage{color}

\usepackage{tikz}
\usetikzlibrary{calc,decorations.markings}
\usepackage{subcaption}

\usepackage{colortbl}
\usepackage{lipsum} 

\usepackage{ragged2e}

\usepackage[pdftex,colorlinks=true,linkcolor=cyan,citecolor=red,urlcolor=blue]{hyperref}

\DeclareMathAlphabet\mathbfcal{OMS}{cmsy}{b}{n}

\begin{document}


\title{On the transfer of the angular momentum of a structured laser pulse to an ensemble of charged particles}
\author{E. Dmitriev}
\author{Ph. Korneev}
 \email{ph.korneev@gmail.com}
\affiliation{ 
National Research Nuclear University MEPhI, Moscow, Russian Federation
}
\affiliation{ 
P.N. Lebedev Physical Institute, Moscow, Russian Federation
}


\begin{abstract}
A structure of a laser pulse may significantly influence the dynamics of interacting particles.
In the case of dilute plasma the particle dynamics may be considered in the single particle approximation.
In this paper the problem of the angular momentum gain of a single particle in a focused structured pulse is considered for some certain cases, including radial and azimuthal polarizations, in the frameworks of the theoretical model, presented in \cite{Dmitriev.Korneev_AngularMomentumGain_PRA-2024}, which includes solving of Maxwell's equations with a required accuracy and the high order perturbation theory.
The obtained analytical results are compared to the results of numerical simulations.
Angular dependence in the structure of the laser pulse is shown to be critically important for the orbital angular momentum gain of an ensemble of charged particles in a structured wave. 
\end{abstract}

\maketitle

\section*{Introduction}

One of the most curious and fundamental problems of interaction of electromagnetic radiation with charged particles is the problem of integrability and existence of integrals of motion.
For known field models, allowing, in addition to other methods, analytical consideration, for instance, for the plane waves or weakly focused Gaussian beams, many results, describing redistribution of momentum and energy between the wave and charged particles, have been obtained.
Methods of nonlinear dynamics, consideration of resonance phenomena and numerical simulations are the robust approaches in these studies.

The energy and momentum in a 3D dynamical system correspond to four additive integrals of motion out of seven, the remaining three correspond to the angular momentum.
However, the problem of the angular momentum transfer in the laser-plasma interactions is studied much less thoroughly, than the problem of the energy and momentum transfer.
It is probably related to a practical complexity of implementations of non-trivial laser pulses configurations with relatively high intensities, i.e. the conditions, when the angular momentum transfer becomes a non-trivial problem. 
The greatest advancements were achieved in the particular cases of axially symmetric pulses with circular polarization.
The angular momentum transfer to the charged particles in this case is known as the inverse Faraday effect \cite{Deschamps1970a,Najmudin2001,Horovitz1997,Tikhonchuk_InverseFaradayEffect1999}. 
One of the most important results is the conclusion that an effective angular momentum transfer to the system of charged particles occurs in the presence of dissipative effects in the system, leading to destruction of the corresponding integrals of motion, see, e.g. \cite{Liseykina2016}.

The problem of the interaction of an electromagnetic wave with a spatial phase and intensity distribution on charged particles does not have a universal solution.
For a single point charge in vacuum, acceleration of the particle may be considered as a consequence of irreversible changes of it's momentum and energy, if the wave sufficiently differs from the long plane wave, where these quantities are restored after the interaction \cite{LandauLifshits_II}. 
During the analysis of the interaction processes of such type, a question of the conservation of the particle's orbital angular momentum could be formulated.
In the case of plasma, when the role of collective motion is significant, redistribution of the orbital angular momentum should influence the structure of the irradiating wave.
As a first step one may, however, limit the problem to the case, when the collective interactions in plasma are negligible, and consider the angular momentum transfer in the single particle approximation. 
The presence of the medium in this approximation reduces to an averaging with the distribution function of the particles.
The required conditions are the smallness of the plasma collective fields with respect to the externally applied, i.e. $e n_e r_D\ll E_0$ where $e$ is the electron charge, $n_e$ is the electron density, $r_D$ is the Debye length in the plasma, $E_0$ is the amplitude of the external field. To stay far from plasma resonances, demand also the diluteness of the plasma, $\omega_p\ll\omega_0$, where $\omega_p=({{4\pi n_e e^2}/{m_e}})^{1/2}$ is the plasma frequency, $m_e$ is the electron mass, $\omega_0$ is the characteristic frequency of the external field, for instance, the main laser wave carrier frequency.
The case of the isotropic plasma distribution is of the primary interest in this context, which, apparently, may be considered as an appropriate initial approximation to real systems.

The problem of the angular momentum transfer to an isotropic system of charged particles is often formulated for fields in the paraxial approximation, which may be described in terms of decomposition in the basis of the Laguerre-Gaussian modes~\cite{Allen1992}. 
For these modes the field structure corresponds to determined values of the orbital angular momentum in the irradiating wave and, strictly speaking, these fields are infinite in time.
The temporal profile may appear to be importnat in the considered problem, as e.g.  a short pulse may be highly asymmetric in terms of it's angular dependence, which, obviously, may significantly facilitate the angular momentum transfer to the particles.
Long pulses, that are implied further are of a fundamental interest; effectiveness of the angular momentum transfer in this case is determined by the absorption rate per unit time.

Numerical simulations with Particle-In-Cell methods \cite{Nuter_PlasmaSolenoid_PRE2018, Nuter_GainOAM_PRE2020}, are probably the first attempts of determining the orbital angular momentum transfer rate from a wave to a system of charged particles in the high intensity regime. The work~\cite{Nuter_PlasmaSolenoid_PRE2018} considers the case of the so-called radial polarization for $a_0=e E_0/(m_e\omega_0c)=1$, where $c$ is the velocity of light in vacuum.
The obtained result turned out to be quite encouraging: the rate of the angular momentum transfer to isotropic plasma appeared to be significant and the effect was considered as one of efficient methods for generation of strongly-magnetized plasma structures with poloidal magnetic fields.
However, details of the process remained rather obscure.
Besides, it remained unclear, what caused the effect to be sufficiently weaker for the same numerical setup for the azimuthal polarization~\cite{private1}.

The analysis of the numerical results faced an issue related to an accurate description of the laser fields with the radial and azimuthal polarizations, as the decomposition in the basis of the Laguerre-Gaussian modes required a large number of terms. Therefore, in \cite{Nuter_PlasmaSolenoid_PRE2018} actually a phenomenological approach was used, when the fields were approximated as the main part (radially polarized) and corrections, which were determined within a certain accuracy using Maxwell equations.
To overcome this, in the subsequent work~\cite{Nuter_GainOAM_PRE2020} certain Laguerre-Gaussian modes were used for the numerical simulations. The observed effect of the orbital angular momentum transfer was of the same order, as that for the radial polarization in \cite{Nuter_PlasmaSolenoid_PRE2018}.
In addition, one of the main and important results obtained in~\cite{Nuter_GainOAM_PRE2020}, was a conclusion that the difference in the phenomenon of the angular momentum gain in the presence and in the absence of the interaction between the particles is only quantitative.
It was also shown, that a longitudinal displacement of interacting particles is vital for the absorption of the angular momentum, however no complete analytical description, explaining the results of numerical calculations, was presented. 
There are several works, where generation of magnetic fields in plasma due to angular momentum transfer was reported. In \cite{Shi2018}, a fluid model of interaction was developed to describe the formation of azimuthal currents in plasma irradiated by two different Laguerre-Gaussian beams. The collective plasma response in plasma was also studies in  \cite{Ali2010,Longman2021,Punia2020}. However, the effect still remained obscure on the basic level, when a structured laser wave interacts with a single particle.

Subsequent attempts of consideration this fundamental problem for a single particle and an ensemble of them involved mainly numerical integration of the equations of motion for the charged particles in prescribed fields~\cite{Tikhonchuk_Numerical_HEDP2020, Toma-2024, Molnar2021}. 
Often those fields were determined with several Laguerre-Gaussian modes, and temporal envelope was factorized.
A more detailed consideration showed that certain attention should be paid to both enough accurate description of the fields and to the particle statistics in an ensemble.
Particularly, the work \cite{Dmitriev_Approximations_BLPI2022} studies the conservation of the integral of motion, based on the conservation of the total angular momentum in the system with circularly polarized pulse and an opposite orbital angular momentum, so that the total (spin and orbital) angular momentum appears to be zero.
The obtained numerical results hinted, that the process of the angular momentum gain is extremely sensitive to the accuracy of the field description and the number of the test particles.
The conclusion that a good statistics is important for an accurate numerical description of the effect also is done in Ref.~\cite{Dmitriev_BrokenAxial_BLPI2023}.
The error from insufficient amount of particles is related to the fact that the gained angular momentum for a single particle appears to be significantly, on the orders of magnitude, higher than that for an isotropic system as a whole~\cite{Toma-2024,Dmitriev.Korneev_AngularMomentumGain_PRA-2024}. 
Within the described background, consistent analytical models, which at least qualitatively could describe the effect of the angular momentum gain in the considered situation, are highly desired.

An attempt to develop such a model was made in  \cite{Dmitriev.Korneev_AngularMomentumGain_PRA-2024}.
In the approximation of moderately intense fields, specifically, when the oscillation amplitude of a charge in the field is smaller than the focused beam waist, which probably may be valid with some accuracy even for $a_0\sim1$, the problem of interaction of an arbitrary light wave with charged particles was considered and the results were applied to the isotropic distribution of the interacting particles.
The general expressions were used to obtain the orbital angular momentum transfer rate for Laguerre-Gaussian modes, calculated with the accuracy, required to obtain a non-zero value.
As an example, in \cite{Dmitriev.Korneev_AngularMomentumGain_PRA-2024} the gained angular momentum was calculated for several low-order Laguerre-Gaussian modes and the combinations of the two of them.
In the present work, less trivial cases are considered based on the general expressions, presented in \cite{Dmitriev.Korneev_AngularMomentumGain_PRA-2024}, in particular, the case of the radial polarization, for which the effect was numerically demonstrated in~\cite{Nuter_PlasmaSolenoid_PRE2018}. 
In addition, the case of the azimuthal polarization and some other illustrative examples are also considered.


\section{Analysis of the orbital angular momentum absorption}



Electromagnetic field of a laser wave may be prescribed with a boundary condition in the focal plane.
Specifying transverse components of the electric fields is sufficient \cite{Thiele2016} to obtain electromagnetic fields in the whole space.
Consider a boundary condition
\begin{equation} \label{eq:bc_rad_azi}
    \mathbf E_\perp \big|_{z = 0} = E_0 g(t) u_{pl}(r, \varphi, z = 0) e^{i \omega_0 t} \mathbf e_\alpha,
\end{equation}
where $g(t)$ is a slow envelope of the electromagnetic wave, $u_{pl}$ is a Laguerre-Gaussian mode~\cite{Allen1992} with the radial index $p$, the azimuthal index $l$ and the beam waist $w_0$, $\varphi$ is the azimuthal angle in the plane, transverse to the laser propagation axis, $r = \sqrt{x^2 + y^2}$,
$\mathbf e_\alpha$ is the unit vector of polarization, equal to the transverse radial unit vector $\mathbf e_0 \equiv \mathbf e_r = \mathbf e_x \cos\varphi + \mathbf e_y \sin\varphi$ if $\alpha = 0$ or the unit azimuthal vector $\mathbf e_1 \equiv \mathbf e_\varphi = - \mathbf e_x \sin\varphi + \mathbf e_y \cos\varphi$ if $\alpha = 1$.

The Laguerre-Gaussian modes~\cite{Allen1992} are defined as

\begin{multline} \label{eq:u_pl}
  u_{pl} \left( r, \varphi, z \right) = C_{pl} \frac{w_0}{w \left( z \right)} \left(\frac{r\sqrt{2}}{w \left( z \right)}\right) ^ {\left| l \right|} \times \\ \times \exp{\left(-\frac{r^{2}}{w^{2} \left( z \right)} \right)} L_{p}^{\left| l \right|}\left(\frac{2 r^{2}}{w^{2} \left( z\right)} \right) \times \\
  \exp {\left( - i l \varphi - \frac{i r^{2} z}{w^2 \left( z\right) z_R}
    + i \left( 2p + \left| l \right| + 1\right) \tan^{-1} \frac{z}{z_R} \right)},
\end{multline}
where $C_{pl} = \sqrt{\frac{2}{\pi} \frac{p !}{\left( p + \left| l \right|\right)!}}$ is the normalization constant, $w(z) = w_0 \sqrt{1 + \frac{z^2}{z_R^2}}$ is the spot size of the beam at the distance $z$ from the focal plane and $z_R = \frac{\pi w_0^2}{\lambda}$ is the Rayleigh range.
The boundary condition $\eqref{eq:bc_rad_azi}$ for the Cartesian components reads
\begin{multline} \label{eq:bc_rad_azi_Cartesian}
    \mathbf E_\perp \big|_{z = 0} = E_0 g(t) u_{pl}(r, \varphi, z = 0) \times \\ \times \text{Re} \left( i^\alpha \left( \mathbf e_x - i \mathbf e_y\right) e^{i \varphi} \right) e^{i\omega_0 t}.
\end{multline}

The electromagnetic field, corresponding to the boundary condition $\eqref{eq:bc_rad_azi}$, may be obtained out of the focal plane $z = 0$ in the frameworks of the paraxial and the slowly varying envelope approximations.

In the frameworks of the slowly varying envelope approximation, $g(t)$ is considered as a slowly varying function compared to the phase term $e^{i\omega_0 t}$, which means that the characteristic pulse duration $\tau$ is considered to be much larger than the field oscillation period: $\tau \gg \omega_0^{-1}$.
Thus, the electromagnetic wave is almost  monochromatic. In the frameworks of the paraxial approximation, the divergence angle of the beam is considered as small, which requires $w_0 \gg c/\omega_0$, a smallness of the characteristic wave length compared to the beam waist.
In this approximation the spatial amplitude may be expressed as a superposition of the Laguerre-Gaussian modes $\eqref{eq:u_pl}$.

As a result, in the used frameworks the solution of the wave equation, corresponding to the boundary condition $\eqref{eq:bc_rad_azi}$ reads
\begin{multline} \label{eq:E_expanded}
    \mathbf E_\perp = E_0 g(t - z/c) \sum \limits_{q = 0}^{\infty} \sum \limits_{m = -\infty}^{\infty} \left( a_{qm \alpha} \mathbf e_x + b_{qm \alpha} \mathbf e_y\right) \times \\ \times u_{qm}(r, \varphi, z) e^{i\omega_0(t - z/c)},
\end{multline}
where $a_{qm\alpha}$ and $b_{qm\alpha}$ are the decomposition coefficients of the Cartesian electric field components.
From the boundary condition $\eqref{eq:bc_rad_azi_Cartesian}$ it follows that $b_{qm\alpha} = (-1)^{1 - \alpha} a_{qm, 1 - \alpha}$.
The longitudinal electric and all the magnetic field components may be obtained from Maxwell equations using the transverse electric field components.
In the frameworks of the approximations used (see, e.g., \cite{Dmitriev.Korneev_AngularMomentumGain_PRA-2024}) the field components read
\begin{equation}
    \begin{aligned}
        H_x & = - E_y, & \ E_z & = - & \frac{c}{\omega_0} \left( i \frac{\partial E_r}{\partial r} + \sum\limits_m \frac{m}{r}E_{m, \varphi} \right), \\
        H_y & = E_x, & \ H_z & = & \frac{c}{\omega_0} \left(i \frac{\partial E_{\varphi}}{\partial r} - \sum\limits_m\frac{m}{r}E_{m, r}\right),
    \end{aligned}
\end{equation}
where $E_{m, r} = E_{m, x} \cos\varphi + E_{m, y} \sin\varphi$, $E_{m, \varphi} = -E_{m, x} \sin\varphi + E_{m, y} \cos\varphi$ and $E_{m, x}$, $E_{m, y}$ are defined as $E_{m, j} = e^{-i m \varphi} \int \frac{d\varphi}{2\pi} E_j e^{i m \varphi}$, where $j = x, y$.
In order to define the decomposition coefficients in $\eqref{eq:E_expanded}$, the orthogonality relation for the Laguerre-Gaussian modes may be used
\begin{equation} \label{eq:orthogonality}
    \int \frac{r dr d\varphi}{w_0^2} u_{pl}(r, \varphi, 0) u_{qm}^*(r, \varphi, 0) = \delta_{pq} \delta_{lm},
\end{equation}
where $\delta_{ij} = 1$, when $i = j$ and $\delta_{ij} = 0$, when $i \neq j$ is the Kronecker symbol and the star stands for complex conjugate.
Then
\begin{equation}
    a_{qm \alpha} = \int \frac{r dr d\varphi}{w_0^2} u_{pl} (r, \varphi, 0) \text{Re}(i^\alpha e^{i\varphi}) u_{qm}^*(r, \varphi, 0).
\end{equation}
After integration over $\varphi$ obtain
\begin{equation}
    a_{qm \alpha} = i^{\alpha} \left( \delta_{m, l - 1} + (-1)^\alpha \delta_{m, l + 1}\right) f_{qm},
\end{equation}
where
\begin{equation} \label{eq:f_qm}
    f_{qm} = \pi \int \frac{r dr}{w_0^2} u_{pl} (r, 0, 0) u_{qm}^* (r, 0, 0).
\end{equation}

It is worth noting, that the easiest case for calculations is $m = l$. According to $\eqref{eq:orthogonality}$, $f_{ql} = \frac{1}{2}\delta_{pq}$.
In the case of $m = l \pm 1$ one can use the explicit form of the Laguerre-Gaussian modes and introduce a variable $y = 2r^2/w_0^2$. Then the expression $\eqref{eq:f_qm}$ reads
\begin{equation} \label{eq:f_qm_int}
    f_{qm} = \frac{\pi C_{pl} C_{qm}}{4} \int \limits_0^\infty y^{\frac{\left| m \right| + \left| l \right|}{2}} e^{-y} L_q^{\left| m \right|} (y) L_p^{\left| l \right|} (y) dy.
\end{equation}

Using the explicit form of the Laguerre polynomials
\begin{equation}
    L_p^{\left| l\right|}(x) = \sum \limits_{k = 0}^p (-1)^k \binom{p + \left| l \right|}{p - k} \frac{x^k}{k!},
\end{equation}
where $\binom{n}{k}$ is the binomial coefficient, one can calculate the integral in $\eqref{eq:f_qm_int}$ as
\begin{multline} \label{eq:f_qm_double_sum}
    f_{qm} = \frac{\pi C_{pl} C_{qm}}{4} \left( q + \left| m\right|\right)! \left( p + \left| l\right|\right)! \times \\
    \times \sum \limits_{k = 0}^{p} \sum \limits_{s = 0}^{q} (-1)^{k + s} \frac{\Gamma\left( k + s + \frac{\left| m\right| + \left| l\right|}{2} + 1\right)}{(q - s)! \left( s + \left| m\right| \right)! (p - k)! \left( k + \left| l \right| \right) k! s!},
\end{multline}
where  $\Gamma(x)$ is the gamma function.

Expressing the gamma function through the generalized binomial coefficient
\begin{equation}
    \binom{x}{k} = \frac{\Gamma\left(x + 1\right)}{\Gamma\left(x - k + 1\right)k!}
\end{equation}
and using the identity (see, e. g., \cite{GradsteinRyzhik})
\begin{equation}
    \sum_{k = 0}^p \binom{n}{k} \binom{m}{p - k} = \binom{n + m}{p},
\end{equation}
one can carry out the summation over $s$. Then $\eqref{eq:f_qm_double_sum}$ reads
\begin{widetext}
\begin{multline}
    f_{qm} = \frac{\pi C_{pl} C_{qm}}{4} \frac{\left( p + \left| l\right|\right)!}{q!} \sum \limits_{k = 0}^{p} (-1)^k \frac{\Gamma\left(k + \frac{\left| m\right| + \left| l\right|}{2} + 1\right)}{(p - k)! (\left| l\right| + k)! k!} \frac{\Gamma\left( \frac{\left| m\right| - \left| l\right|}{2} + q - k\right)}{\Gamma\left( \frac{\left| m\right| - \left| l\right|}{2} - k\right)} = \\
    = \frac{\sqrt{\pi} C_{qm}}{2 C_{pl}} \frac{\Gamma\left( q + \frac{\left| m\right| - \left| l\right|}{2}\right) \Gamma\left( \frac{\left| m\right| + \left| l\right|}{2} + 1\right)}{ \left| l\right|! q! \Gamma\left( \frac{\left| m\right| - \left| l\right|}{2}\right)} \times \\
    \times {}_3 F_2 \left( 1 - \frac{\left| m\right| + \left| l\right|}{2}, 1 - \frac{\left| m\right| - \left| l\right|}{2}, -p; 1 + \left| l \right|, 1 - \frac{\left| m\right| - \left| l\right|}{2} - q; 1\right),
\end{multline}
where
\begin{equation}
    {}_3 F_2 (\alpha_1, \alpha_2, \alpha_3; \beta_1, \beta_2; z) = \sum \limits_{k = 0}^{\infty} \frac{(\alpha_1)_k (\alpha_2)_k (\alpha_3)_k}{(\beta_1)_k (\beta_2)_k} \frac{z^k}{k!}
\end{equation}
is the generalized hypergeometric series and $(\alpha)_k = \alpha (\alpha + 1) \dots (\alpha + k - 1)$.
With the coefficients $f_{qm}$, one can calculate the Cartesian and the cylindrical components of the electric field, then
\begin{multline} \label{eq:E_rad_azi}
    \mathbf E_\perp = E_0 g(t - z/c) i^\alpha \sum \limits_{q = 0}^{\infty} \left\{ \mathbf e_x \left( f_{q, l - 1} u_{q, l - 1} + (-1)^\alpha f_{q, l + 1} u_{q, l + 1} \right) - \right.\\
    \left. - i \mathbf e_y \left( f_{q, l - 1} u_{q, l - 1} - (-1)^\alpha f_{q, l + 1} u_{q, l + 1} \right) \right\} e^{i\omega_0(t - z/c)} = \\
    = E_0 g(t - z/c) i^\alpha \sum \limits_{q = 0}^{\infty} \left\{ \mathbf e_r \left( f_{q, l - 1} u_{q, l - 1} e^{-i \varphi} + (-1)^\alpha f_{q, l + 1} u_{q, l + 1} e^{i\varphi}\right) - \right.\\
    \left. - i \mathbf e_\varphi \left( f_{q, l - 1} u_{q, l - 1} e^{-i \varphi} - (-1)^\alpha f_{q, l + 1} u_{q, l + 1} e^{i\varphi}\right) \right\} e^{i\omega_0(t - z/c)}.
\end{multline}

\end{widetext}

One can see that both the main (radial or azimuthal) and the additional (azimuthal or radial respectively) components are present in the expansion \eqref{eq:E_rad_azi}.
It is worth noting that the additional component is present also in the focal plane, which is caused by the incorrect asymptotics of the boundary condition $\eqref{eq:bc_rad_azi}$ near the laser axis.
The $m$-th decomposition component of the Cartesian component of an exact solution of the wave equation in the paraxial approximation (which depends on $\varphi$ as $\sim e^{-i m \varphi}$) should approach $r \to 0$ as $\sim r^{\left| m \right|}$, however it is not so in the case of the boundary condition $\eqref{eq:bc_rad_azi}$.
The incorrect asymptotics of the boundary condition leads to the difference of the main component in $\eqref{eq:E_rad_azi}$ from the boundary condition $\eqref{eq:bc_rad_azi}$ in the focal plane.
The terms in $\eqref{eq:E_rad_azi}$ have the correct asymptotics, and the whole expression $\eqref{eq:E_rad_azi}$ is a solution of the wave equation in the paraxial approximation in the case of $g(t) = 1$.
Note also, that the radial and azimuthal components of electric fields depend on $\varphi$ as $E_r, E_\varphi \sim e^{-il\varphi}$.

As mentioned above, there is no exact solution of the Maxwell's equations in the form of a wave, traveling along the $z$ axis, which coincides in the focal plane with the boundary condition $\eqref{eq:bc_rad_azi}$.
However, using this boundary condition, one may use the method presented in Ref.~\cite{Thiele2016} and obtain an exact analytical solution. This method is implemented in several numerical codes, solving Maxwell equations, as, e.g. in~\cite{Derouillat2018}. Also this method was used to define the radially polarized pulses in~\cite{Nuter_PlasmaSolenoid_PRE2018}.
The obtained solution does not exactly satisfy the boundary condition, however it is an exact solution of the Maxwell equations and is approximately equal to the boundary condition in the focal plane far enough from the laser pulse propagation axis.
In addition, in the frameworks of the used paraxial and slowly varying envelope approximations it will coincide with the approximate solution $\eqref{eq:E_rad_azi}$, see \cite{Thiele2016, Gonzalez2018}.
The exact solution reads
\begin{equation} \label{eq:E_exact}
    \mathbf E_\perp = \int \limits_{\omega^2/c^2 - k_\perp^2 > 0} \frac{d\omega d\mathbf k_\perp}{(2 \pi)^3} \mathbf E_\perp(\omega, \mathbf k_\perp, 0) e^{i \left(\omega t - \mathbf k_\perp \mathbf r_\perp - k_z z\right)},
\end{equation}
where $k_z = \frac{\omega}{c} \sqrt{1 - \frac{k_\perp^2 c^2}{\omega^2}}$ and
\begin{equation} \label{eq:E_omega_k}
    \mathbf E_\perp(\omega, \mathbf k_\perp, 0) = \int d t d\mathbf r_\perp \mathbf E_\perp \Big|_{z = 0} e^{- i \left(\omega t - \mathbf k_\perp \mathbf r_\perp\right)}.
\end{equation}
The remaining components of the electromagnetic field may be calculated using the transverse components of the electric field with Maxwell equations.
Calculation of the integral $\eqref{eq:E_omega_k}$ leads to
\begin{widetext}
\begin{multline} \label{eq:E_int_phi}
    \mathbf E_\perp(\omega, \mathbf k_\perp, 0) = E_0 i^\alpha \pi g_{\omega - \omega_0} e^{-i l \theta} \left\{ \mathbf e_x \left( e^{i \theta} i^{|l - 1|} h_{l - 1}(k_\perp) + (-1)^\alpha e^{-i \theta} i^{|l + 1|} h_{l + 1}(k_\perp)\right) - \right.\\
    \left.- i \mathbf e_y \left( e^{i \theta} i^{|l - 1|} h_{l - 1}(k_\perp) - (-1)^\alpha e^{-i \theta} i^{|l + 1|} h_{l + 1}(k_\perp)\right)\right\},
\end{multline}
where $g_\omega = \int \limits_{-\infty}^\infty dt g(t) e^{-i\omega t}$, $h_{m}(k_\perp) = \int \limits_0^\infty u_{pl}(r, 0, 0) J_{|m|}(k_\perp r) r dr$ and $\theta$ denotes the angle between the vector $\mathbf k_\perp$ and the $x$ axis, and the formula
\begin{equation}
  \int \limits_0^{2\pi} d\varphi \exp\left(i s \varphi + i z \cos\varphi\right) = 2 \pi i^{|s|}J_{|s|}(z)
\end{equation}
was used.
Inserting $\eqref{eq:E_int_phi}$ to $\eqref{eq:E_exact}$ and integration over $\theta$ gives
\begin{multline}
    \mathbf E_\perp = E_0 i^{\alpha} e^{- i l \varphi} \int \limits_{\omega^2/c^2 - k_\perp^2 > 0} \frac{d\omega k_\perp dk_\perp}{4 \pi} g_{\omega -\omega_0} \left\{\mathbf e_x \left( e^{i \varphi}F_{l - 1}(k_\perp, r) + (-1)^\alpha e^{-i \varphi} F_{l + 1}(k_\perp, r) \right) - \right.\\
    \left.- i \mathbf e_y \left( e^{i\varphi}F_{l - 1}(k_\perp, r) - (-1)^\alpha e^{-i \varphi} F_{l + 1}(k_\perp, r) \right)\right\} e^{i (\omega t- k_z z)},
\end{multline}
where $F_{m}(k_\perp, r) = h_{m}(k_\perp) J_{|m|}(k_\perp r)$.
Then, introducing
\begin{equation}
    \mathcal{F}_m (r, z, t) = \int \limits_{\omega^2/c^2 - k_\perp^2 > 0} \frac{d\omega k_\perp dk_\perp}{4 \pi} g_{\omega -\omega_0} F_m(k_\perp, r) e^{i(\omega t - k_z z)},
\end{equation}
one obtains the exact solution in the form
\begin{multline}
\label{eq:exact}
    \mathbf E_\perp = E_0 i^\alpha e^{-i l \varphi} \left\{ \mathbf e_x \left( \mathcal{F}_{l - 1}e^{i\varphi} + (-1)^\alpha \mathcal{F}_{l + 1}e^{-i\varphi}\right) - i \mathbf e_y \left( \mathcal{F}_{l - 1}e^{i\varphi} - (-1)^\alpha \mathcal{F}_{l + 1}e^{-i\varphi}\right)\right\} = \\
    = E_0 i^\alpha e^{-i l \varphi} \left\{ \mathbf e_r \left( \mathcal{F}_{l - 1} + (-1)^\alpha \mathcal{F}_{l + 1}\right) - i \mathbf e_\varphi \left( \mathcal{F}_{l - 1} - (-1)^\alpha \mathcal{F}_{l + 1} \right)\right\}.
\end{multline}
\end{widetext}

As it follows from \eqref{eq:exact}, the radial and the azimuthal components depend on $\varphi$ as $\sim e^{-i l \varphi}$.
Similarly one can show that other cylindrical components of the electromagnetic field depend on $\varphi$ as $\sim e^{- i l \varphi}$ as well.
So, the components of the exact solution of Maxwell equations with boundary condition for both radial and azimuthal polarizations $\eqref{eq:bc_rad_azi}$, obtained with the method, described in the paper \cite{Thiele2016}, depend on $\varphi$ as $E_r, E_\varphi, E_z, H_r, H_\varphi, H_z \sim e^{- i l \varphi}$.
\onecolumngrid
\begin{center}
\begin{figure}[H]
            \includegraphics[width=\linewidth]{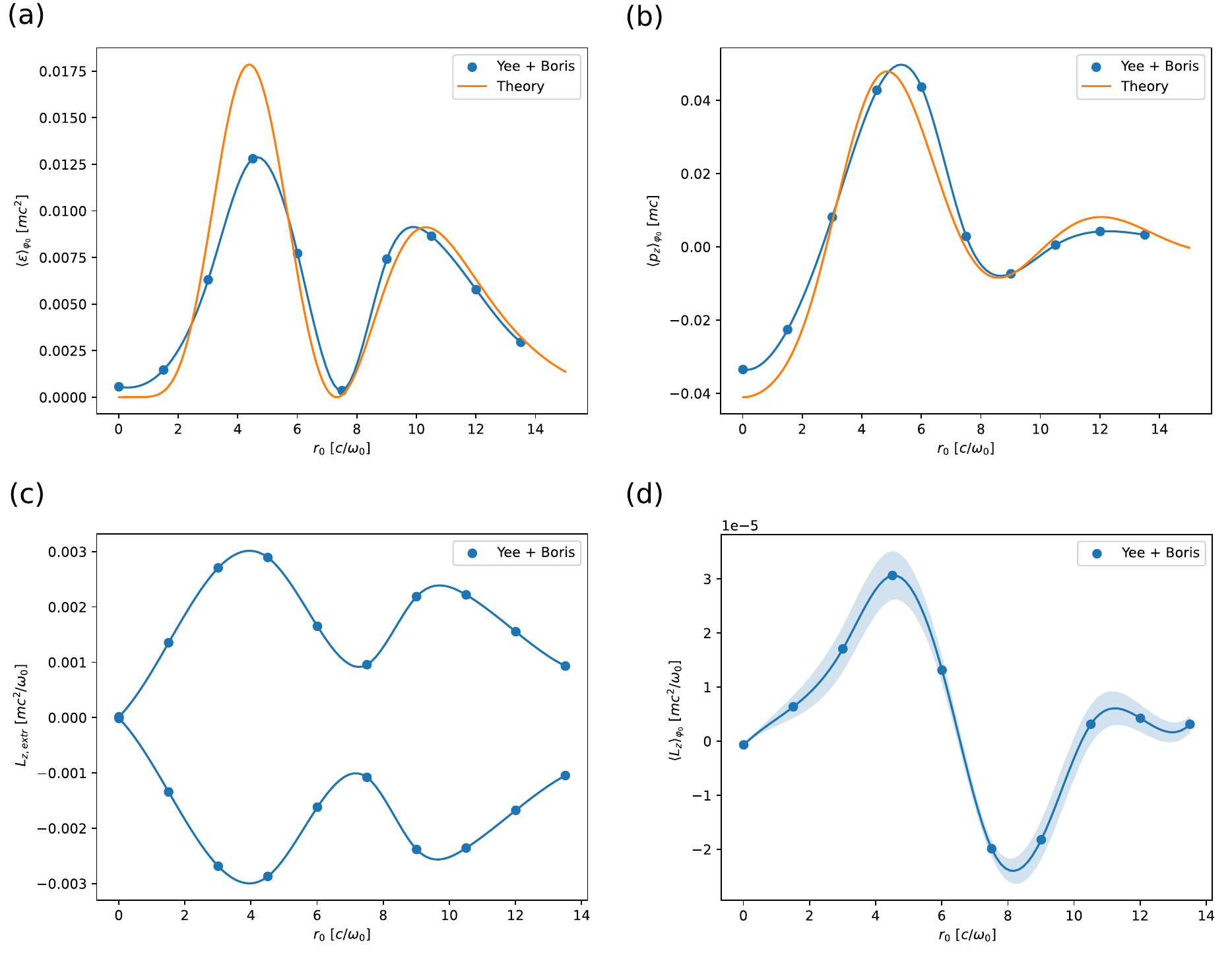}
                \caption{Dependences of (a) average gained energy, (b) average gained longitudinal momentum, (c) extreme values of gained angular momentum and (d) average gained angular momentum of electrons on the initial distance from the laser axis after the interaction with the laser pulse. Blue dots represent numerical simulation results, connected with interpolation. Orange lines represent theoretical dependence. Radial polarization case. \vspace{0.5cm}}
            \label{fig:rad}
\end{figure} 
\end{center}
\twocolumngrid

Far from the laser pulse propagation axis, the expression for the main component may be simplified, if one assumes that the sum in $\eqref{eq:E_rad_azi}$ can be evaluated approximately, by considering $l \pm 1 \approx l$. 
Then
\begin{equation} \label{eq:approx_calculation}
    \sum \limits_{q = 0}^\infty \left( f_{q, l - 1} u_{q, l - 1} + f_{q, l + 1} u_{q, l + 1} \right) \approx 2 \sum \limits_{q = 0}^\infty f_{ql} u_{ql} = u_{pl},
\end{equation}
where $f_{ql} = \frac{1}{2}\delta_{pq}$ was used.
Then, the main component far from the laser pulse propagation axis may be roughly approximated as
\begin{equation} \label{eq:E_beta_approx}
    E_\alpha \ \approx \ E_0 g(t - z/c) u_{pl}(r, \varphi, z) e^{i\omega_0(t - z/c)},
\end{equation}
where $E_\alpha$ means $E_0 \equiv E_r$ when $\alpha = 0$ (the radially polarized case) and $E_1 \equiv E_\varphi$ when $\alpha = 1$ (the azimuthally polarized case).

The approximation $\eqref{eq:E_beta_approx}$ was used in the paper \cite{Nuter_PlasmaSolenoid_PRE2018} as a model expression for the radial component of the electric field, where an agreement was demonstrated between the numerical and analytical results. 
As it was described above, this approximate expression may be up to some extent valid far from the laser pulse axis, where the most efficient interaction of the laser pulse with the charged particles is expected.
However, a more consistent description of the interaction of a charged particles with the electromagnetic wave, considered in the paper \cite{Nuter_PlasmaSolenoid_PRE2018}, requires use of the fields $\eqref{eq:E_rad_azi}$. 
\onecolumngrid
\begin{center}
\begin{figure}[h]
            \includegraphics[width=\linewidth]{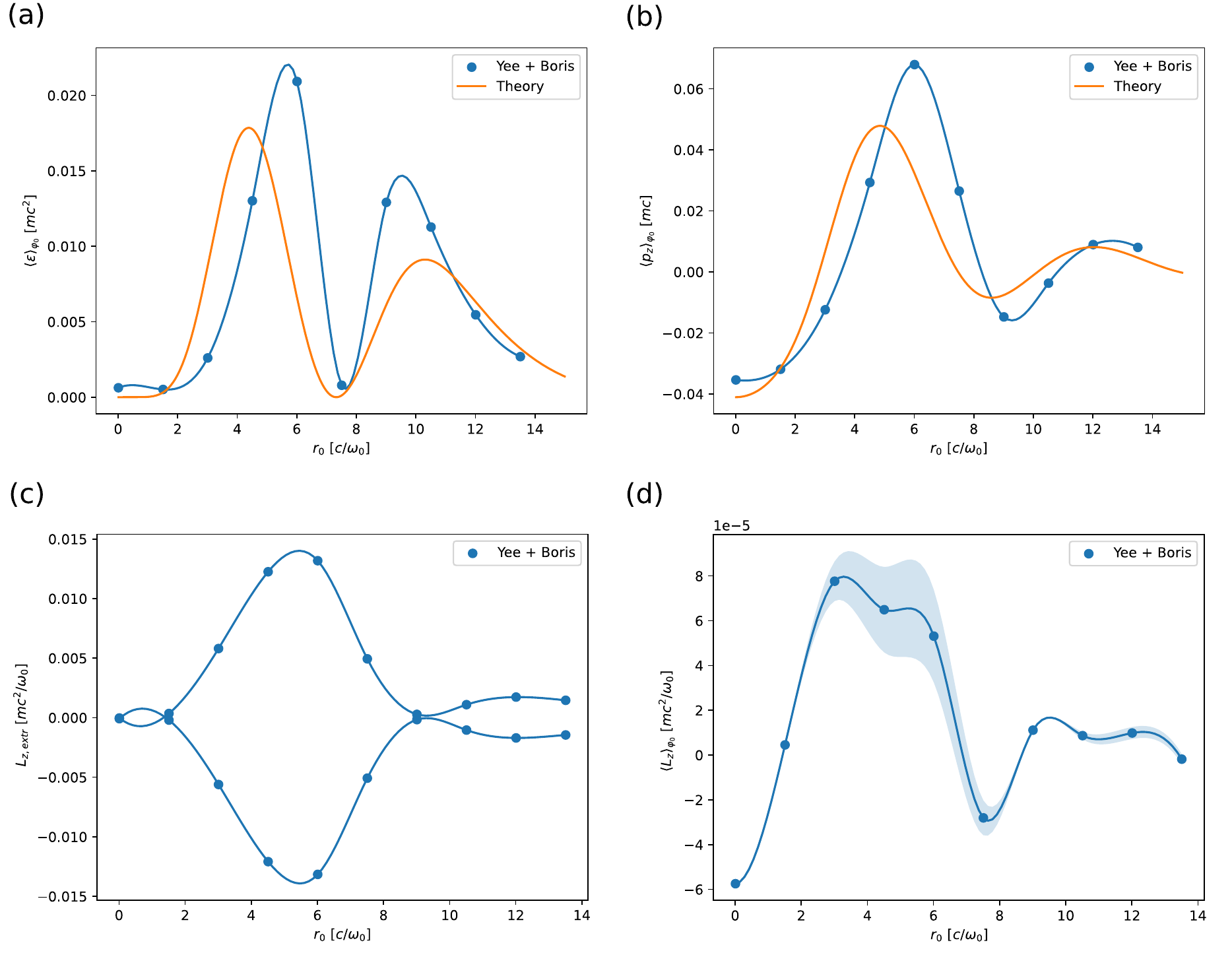}
                \caption{\justifying Dependences of (a) average gained energy, (b) average gained longitudinal momentum, (c) extreme values of gained angular momentum and (d) average gained angular momentum of electrons on the initial distance from the laser axis after the interaction with the laser pulse. Blue dots represent numerical simulation results, connected with interpolation. Orange lines represent theoretical dependence. Azimuthal polarization case.}
            \label{fig:azi}
\end{figure} 
\end{center}
\twocolumngrid
As it was demonstrated in \cite{Dmitriev.Korneev_AngularMomentumGain_PRA-2024}, the angular momentum gain by the charged particles may be described in the frameworks of the perturbation theory with respect to the small parameter of the dimensionless field amplitude $a_0 \ll 1$, and the perturbation theory may be used for estimates even at $a_0 \sim 1$ for reasonably focused pulses with $w_0\gtrsim c/\omega_0$.
According to the results of the perturbation theory, the angular momentum gain by a system of isotropically distributed particles occurs as an effect of the fourth order with respect to $a_0$.
Then, an electron gains energy and longitudinal momentum after the interaction with the wave $\eqref{eq:E_rad_azi}$ according to the following estimates
\begin{equation} \label{eq:ekin_pz_theor}
  \begin{aligned}
    \varepsilon & \approx m_e c^4 \frac{a_0^4}{32} \left( \frac{\partial}{\partial r_0} \left( \left| \mathcal{E}_0\right|^2 + \left| \mathcal{E}_1\right|^2 \right)\right)^2 \tau_{int}^2, \\
    p_z & \approx \frac{\varepsilon}{c} - m_e c^2 \frac{a_0^2}{4} \frac{\partial}{\partial z_0} \left( \left| \mathcal{E}_0\right|^2 + \left| \mathcal{E}_1\right|^2 \right) \tau_{int}, \\
  \end{aligned}
\end{equation}
where $\tau_{int} = \int \limits_{-\infty}^\infty g(t) dt$ is the characteristic time of interaction, the lower index $0$ stands for the initial coordinate of the electron and
\begin{multline}
      \mathcal{E}_\beta \equiv  f_{q, l - 1} u_{q, l - 1} (r_0, \varphi_0, z_0) e^{-i \varphi_0} +  \\ +(-1)^\beta f_{q, l + 1} u_{q, l + 1} (r_0, \varphi_0, z_0) e^{i\varphi_0},
\end{multline}
describes the spatial amplitudes of the components of the electromagnetic wave.
According to the results, presented in \cite{Dmitriev.Korneev_AngularMomentumGain_PRA-2024}, the gained angular momentum may be estimated in the second order of the perturbation theory on $a_0$ as
\begin{equation}
    L_z^{(2)} \approx - m_e c^2 \frac{a_0^2}{4} \frac{\partial}{\partial \varphi_0} \left( \left| \mathcal{E}_0\right|^2 + \left| \mathcal{E}_1\right|^2 \right) \tau_{int} = 0,
\end{equation}
As long as $\left| \mathcal{E}_\beta\right|^2$ does not depend on $\varphi_0$ for $\beta = 0, \ 1$, the gained angular momentum in the second order turns to zero.
The first non-vanishing contribution to the average gained angular momentum appears in the fourth order perturbation theory. It reads $\langle L_z^{(4)} \rangle_{\varphi_0} = \langle \left( \mathbf r^{(2)} \cdot \nabla_0\right) L_z^{(2)} \rangle_{\varphi_0}$, where $\mathbf r^{(2)}$ is the displacement of the particle in the second order of the perturbation theory.
For the considered radial and azimuthal polarizations, this first non-vanishing for some other fields value for the average gained angular momentum also turns to zero.
Probably, to obtain a non-zero result, next orders of the perturbation theory should be considered.
\onecolumngrid
\begin{center}
\begin{figure}[H]
            \includegraphics[width=\linewidth]{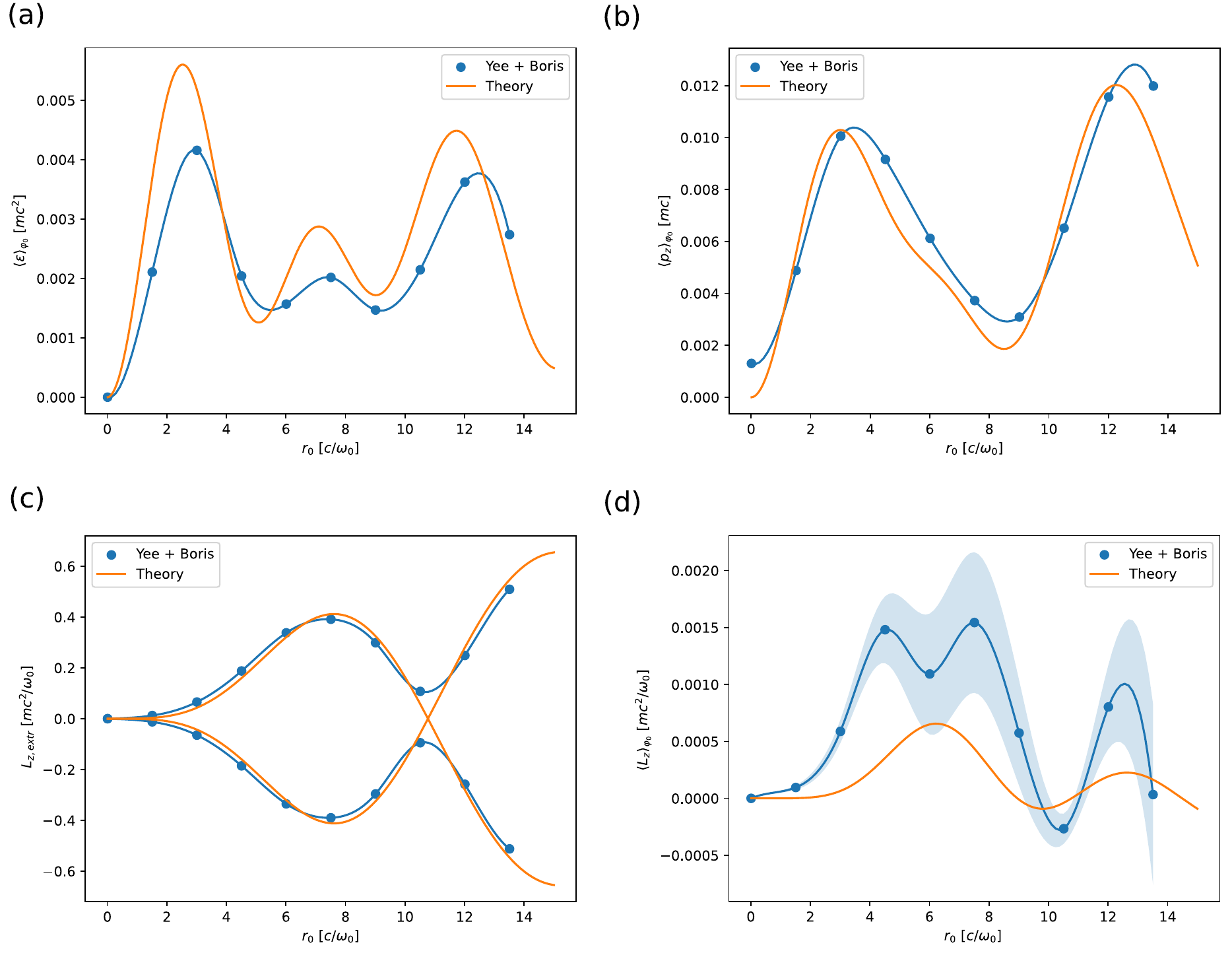}
                \caption{Dependences of (a) average gained energy, (b) average gained longitudinal momentum, (c) extreme values of gained angular momentum and (d) average gained angular momentum of electrons on the initial distance from the laser axis after the interaction with the laser pulse. Blue dots represent numerical simulation results, connected with interpolation. Orange lines represent theoretical dependence. Superposition case $p=1,l=1$ and $q=0,m=3$.}
            \label{fig:comb}
\end{figure} 
\end{center}
\twocolumngrid
Compare the findings of the approximate analytical model with accurate numerical modeling. For that, consider the interaction of an ensemble of isotropically distributed charged particles with an electromagnetic wave with the same parameters, using the open-source Particle-In-Cell code Smilei \cite{Derouillat2018}, where the fields are defined with the algorythm, described above.
In the simulations, the interactions between particles are switched off.
This setup allows to numerically calculate the fields, acting on the particles with the Yee scheme, and to move the particles with the Boris-pusher procedure.
In the numerical simulations, a cylindrical computational grid with $6000$ cells in the transverse direction and $1600$ cells in the longitudinal direction with a spatial resolution of $2.5$ nm is considered.
The laser wave propagates from the left boundary $z = 0$, where the boundary condition
\begin{equation}\label{eq:bcPIC}
    \mathbf H_\perp \big|_{z = 0} = (-1)^\alpha E_0 g(t) u_{pl}(r, \varphi, z = 0) e^{i \omega_0 t} \mathbf e_{1 - \alpha}
\end{equation}
is set.
Within the approximation of the slowly varying temporal envelope and the paraxial approximation, the boundary condition \eqref{eq:bcPIC} corresponds to the radially or azimuthally polarized electric field, prescribed by the boundary condition $\eqref{eq:bc_rad_azi}$, when $\alpha = 0$ or $\alpha = 1$ correspondingly.
Physically, the carrier laser frequency is $\omega_0 = 2.3 \times 10^{15} \ s^{-1}$, the beam waist in the focal plane is $w_0 = 1.3~\mu$m and duration of the laser pulse is $\tau = 2 \pi n / \omega_0$, where $n = 6$ is the number of wave periods.
The dimensionless amplitude of the wave is $a_0 = 1$, the radial and azimuthal indexes are correspondingly $p = 0$ and $l = 1$.
The temporal envelope of the laser pulse is $g(t) = \cos^2\left(\frac{t - \tau/2}{\tau} \pi\right)$, when $\left|t - \tau / 2\right| < \tau/2$ and $0$ otherwise.
On the boundaries, the absorbing boundary conditions for electromagnetic fields are considered.
For a better statistics, electrons are placed on the distance $2.5~\mu$m from the $z = 0$ plane at given distances from the laser pulse axis at random values of initial azimuthal angles.
The initial radia were considered to be $10$ equally distributed values of from $r_0 = 0$ to $r_0 = 1.9~\mu$m, the particles were initialized on the rings with axes, coinciding with the laser pulse axis.

Dependencies of the particle average energy, average longitudinal momentum, average angular momentum and the extreme values of the angular momentum, gained by the particles in the radially and azimuthally polarized fields are shown in Fig. \ref{fig:rad} and \ref{fig:azi} respectively.
The averaging is performed with respect to the initial angle $\varphi_0$.
The blue dots represent the results of the numerical simulations, between the points the dependencies are obtained with the cubic interpolation.
The orange lines show the predictions of the perturbation theory, according to the expressions $\eqref{eq:ekin_pz_theor}$, for the orbital angular momentum the predictied value is zero and not showed.
The figures with the average gained angular momentum also show the region of the statistical errors, arising due to the averaging, which may be estimated as
\begin{equation}
    \Delta L_z = \frac{\sqrt{\langle \left(L_z - \langle L_z \rangle_{\varphi_0} \right)^2 \rangle_{\varphi_0} }}{\sqrt{N}},
\end{equation}
where $N$ is the number of particles, used in the calculation of $\langle L_z \rangle_{\varphi_0}$ at given $r_0$.

As it follows from Figs. \ref{fig:rad} and \ref{fig:azi}, the energy and the longitudinal momentum, gained in the numerical modeling, qualitatively coincide with the theoretical values, and that the average gained angular momentum is several orders smaller, than the extreme values of the gained angular momentum.
It is worth noting that from the obtained numerical results, the observed gain of the energy, momentum and orbital angular momentum is more effective in the azimuthally polarized field, than in the radially polarized field.

Additionally, a case of the superposition of two linearly polarized Laguerre-Gaussian modes with different combinations of the indexes is considered. The boundary condition for the field is chosen as
\begin{multline} \label{eq:bc_comb}
    \mathbf H_\perp \big|_{z = 0} = E_0 g(t) \times \\ \times \frac{u_{pl}(r, \varphi, z = 0) + u_{qm}(r, \varphi, z = 0)}{\sqrt{2}} e^{i \omega_0 t} \mathbf e_x.
\end{multline}
In the numerical modeling the same parameters are used as for the cases of the radial and azimuthal polarizations above, except for the indexes, which are $p = 1$, $l = 1$ and $q = 0$, $m = 3$.
The results of the numerical modeling and their comparison with the theoretical predictions are presented in Fig. \ref{fig:comb}.

Alongside with the nonzero energy and the longitudinal momentum gain, the perturbation theory predicts nonzero average values of the gained orbital angular momentum according to the expressions from Ref. \cite{Dmitriev.Korneev_AngularMomentumGain_PRA-2024}, which read
\begin{widetext}

\begin{equation} \label{eq:theor_comb}
  \begin{aligned}
    \langle \varepsilon \rangle_{\varphi_0} \approx & m_e c^4 \frac{a_0^4}{128} \left( \left( \frac{\partial |u_{pl}|^2}{\partial r_0} + \frac{\partial |u_{qm}|^2}{\partial r_0}\right)^2 + 2 \Big|\frac{\partial \left( u_{pl} u_{qm}^*\right)}{\partial r_0} \Big|^2\right)\Big|_{\varphi_0 = 0} \tau_{int}^2, \\
    \langle p_z \rangle_{\varphi_0} \approx & - m_e c^2 \frac{a_0^2}{8} \frac{\partial \left( |u_{pl}|^2 + |u_{qm}|^2\right)}{\partial z_0} \Big|_{\varphi_0 = 0} \tau_{int}. \\
    L_{z, extr} \approx & \pm m_e c^2 \frac{a_0^2}{4} (l - m) \big| u_{pl} u_{qm} \big| \Big|_{\varphi_0 = 0} \tau_{int}, \\
    \langle L_z \rangle_{\varphi_0} \approx & - \frac{m_e c^4}{\omega_0} (l - m) \frac{a_0^4}{64} \text{Re} \left\{ i \left( \frac{\partial A}{\partial r_0} \frac{\partial B^*}{\partial r_0} + \frac{(l - m)^2}{r_0^2} A B^* + \frac{2 \omega_0}{c} A \frac{\partial A^*}{\partial z_0}\right)\right\} \tau_{int}^2, \\
    \end{aligned}
\end{equation}
\end{widetext}
where the functions $A$ and $B$ are defined for the given Laguerre-Gaussian modes as $A = u_{qm}^* u_{pl} \Big|_{\varphi_0 = 0}$ and $B = \frac{i z_0}{2} \left( u_{qm}^* \frac{\partial u_{pl}}{\partial z_0} - u_{pl} \frac{\partial u_{qm}^*}{\partial z_0} \right) \Big|_{\varphi_0 = 0}$.

The dependencies of the gained energy, longitudinal momentum and angular momentum, obtained with the numerical modeling, are in a good quantitative agreement with the theoretical ones, while the numerical dependence of the average gained orbital angular momentum is qualitatively similar to the theoretical dependence.
It is worth noting, that the relative statistical errors in this case appear to be greater in general, than in the case of radial and azimuthal polarizations. 
Probably, it is due to the fact that the errors are determined by the order of magnitude of the extreme values of the gained orbital angular momentum.
The difference between the extreme values and the average values is greater in the case of the superposition, than in the cases of the radial or azimuthal polarizations, which leads to the greater relative errors.

The substantial difference between the cases of radial and azimuthal polarization and the case of the superposition of Laguerre-Gaussian modes is the presence of the azimuthal angle dependence of the intensity of the electromagnetic wave, appearing in the developed perturbation theory.
In addition, the scales of the gained angular momentum in the cases of radially and azimuthally polarized fields differ from the scales in the case of the superposition by several orders, which is, apparently, related to the angular dependencies of the fields intensity.

\section{Conclusion}

Based on the analysis of the obtained results, one can conclude, that the orbital angular momentum gain by the charged particles, interacting with a structured wave appears to be suppressed in the absence of the explicit angular dependence in the field intensity.
At the same time, the results following from the analytical expressions, obtained in the work \cite{Dmitriev.Korneev_AngularMomentumGain_PRA-2024} are different from the previously obtained numerical modeling results in the literature, in particular for the radial field. 
Interestingly, the modeling, carried out in the present work, shows an intermediate result between the previously obtained significant values of the gained orbital angular momentum and the analytical predictions in the frameworks of the considered 4-th order perturbation theory.
On the contrary, the carried out analysis of the angular momentum gain for the cases of Laguerre-Gaussian beams shows the qualitative agreement between the numerical results and the results of the analytical consideration, where the discrepancy may be a consequence of a high field amplitude values, which are marginal for the used perturbation theory.

Still, the understanding of the process of the interaction between a structured light beam with the orbital angular momentum and particles on the basic level is yet not fully clear. It may appear, for instance, that the numerical consideration requires some additional benchmarking and development or that the perturbation theory analysis is missing some important features of the interaction, like e.g. stochastic processes or presence of dynamical resonances of different sorts.


\section{ACKNOWLEDGMENTS}


The work was funded by the Russian Science Foundation
under Grant No. 24-22-00402. The authors acknowledge the NRNU MEPhI High-
Performance Computing Center and the Joint Supercomputer Center of RAS.

\end{document}